\documentclass[12pt,preprint]{aastex}
\usepackage{natbib}
\usepackage{graphicx}
\shorttitle{The Existence of Sterile Neutrino Halos in Galactic Centers}
\begin{document}
\title{The Existence of Sterile Neutrino Halos in Galactic Centers as an 
Explanation of the Black Hole mass - Velocity Dispersion Relation}
\author{M.~H.~Chan and M.~-C.~Chu}
\affil{Department of Physics and Institute of Theoretical Physics,
\\ The Chinese University of Hong Kong,
\\  Shatin, New Territories, Hong Kong, China}
\email{mhchan@phy.cuhk.edu.hk, mcchu@phy.cuhk.edu.hk}

\begin{abstract}
If sterile neutrinos exist and form halos in galactic centers, they can 
give rise to observational consequences. In particular, the sterile 
neutrinos decay radiatively and heat up the gas in the protogalaxy to 
achieve hydrostatic equilibrium, and they provide the mass to form 
supermassive blackholes. A natural correlation between the blackhole mass 
and velocity dispersion thus arises $\log(M_{BH,f}/M_{\odot})=\alpha \log 
(\sigma /200 
{\rm km~s^{-1}})+ \beta$ with 
$\alpha \approx 4$ and $\beta \approx 8$. 
\end{abstract} 

\section{Introduction}
Understanding the nature of dark matter remains a fundamental problem in
astrophysics and cosmology. Since the discovery of neutrinos' non-zero
rest mass \citep{Fukuda, Bilenky}, the possibility that neutrinos
contribute to cosmological dark matter has become a hot topic again. In
particular, the sterile neutrinos are a class of candidate dark matter
particles with no standard model interaction. Although the recent
MiniBooNE 
result challenges the LSND result that suggests the existence of eV scale 
sterile neutrinos \citep{Mini}, more massive sterile 
neutrinos (eg.~keV) may still exist. The fact that active 
neutrinos have rest mass implies that right-handed neutrinos should 
exist which may indeed be massive sterile neutrinos with rest mass $m_s$. 

Recently, it was proposed that a degenerate sterile neutrino halo exists 
in the galactic center \citep{Viollier,Munyaneza2}. Later, this model 
is
ruled out with the availability of more precise data, particularly those 
on orbits of stars such as S2 near the Milky Way
center \citep{Schodel}. Nevertheless, the existence of such a sterile
neutrino halo at the centers of protogalaxies may still be possible. 
Sterile neutrinos
decay into lighter neutrinos and photons, and provide mass to 
fuel the growth of supermassive
blackholes. However, since most of these sterile neutrinos may have either 
decayed or fallen into the supermassive blackhole, the total mass of the 
sterile neutrino halo at the galactic center becomes
very small ($\ll 10^6 M_{\odot}$) at present. Moreover, the 
existence of decaying sterile neutrinos may 
help to solve the cooling flow problem in clusters \citep{Chan} as well as
reionization in the 
universe \citep{Hansen}. Therefore it is worthwhile to discuss 
the consequences of the existence of massive sterile neutrinos at the
centers of protogalaxies, which decay into light neutrinos and photons.

On the other hand, recent observations have led to some tight relations 
between the central 
black 
hole masses $M_{BH,f}$ and velocity dispersions $\sigma$ in the bulges of 
galaxies. These 
relations 
can be summarized as $\log(M_{BH,f}/M_{\odot})= \alpha \log(\sigma /200 
{\rm 
~km~s^{-1}})+ \beta$, where $\alpha$ is found to be $3.75 \pm 0.3$ 
\citep{Gebhardt} and $4.8 \pm 0.5$ \citep{Ferrarese} 
by two different groups. \citet{Tremaine} reanalysed both sets of 
data and obtained $\alpha=4.02 \pm 0.32$ and 
$\beta=8.13 \pm 0.06$. These results indicate that 
blackhole formation may be related to galaxy formation, which 
challenges existing galaxy formation theories \citep{Adams}. 

This relation has been derived in recent theoretical models 
\citep{Adams,MacMillan,Robertson,Murray,King,King2}. We assume that 
a degenerate 
sterile neutrino halo exists in the center of a protogalaxy. There are two
different decay modes for sterile neutrinos $\nu_s$. The major decaying
channel is $\nu_s \rightarrow 3 \nu$ with decay rate 
\citep{Barger,Boyarsky}
\begin{equation}
\Gamma_{3 \nu}= \frac{G_F^2}{384 \pi ^3} \sin^22 \theta m_s^5=1.77 \times
10^{-20} \sin^22 \theta \left( \frac{m_s}{1~\rm keV} \right)^5~{\rm 
s^{-1}},
\end{equation}
where $G_F$ and $\theta$ are the Fermi constant and mixing angle of
sterile neutrino with active neutrinos respectively. The minor decaying
channel is $\nu_s \rightarrow \nu + \gamma$ with decay rate 
\citep{Barger,Boyarsky}
\begin{equation}
\Gamma= \frac{9 \alpha G_F^2}{1024 \pi ^4} \sin^22 \theta m_s^5=1.38
\times 10^{-22} \sin^22 \theta \left( \frac{m_s}{1~ \rm keV} 
\right)^5~{\rm s^{-1}}
\approx \frac{1}{128} \Gamma_{3 \nu},
\end{equation}
where $\alpha$ is the fine structure constant. It is quite difficult to
detect the active neutrinos produced in the major decaying channel.
Therefore, we focus on observational consequence of the radiative decay of
the sterile neutrinos with rest mass $m_s \ge 10$ keV. They emit 
high energy photons ($\approx m_s/2$) which heat up the 
surrounding gas so that hydrostatic equilibrium of the latter is 
maintained. The 
sterile neutrino halo also provides mass to form a supermassive blackhole 
from a small seed blackhole. Without any further assumption, $\alpha 
\approx 4$ is consistent with the range of the decay rate obtained by 
observational data from cooling flow clusters. 
In this article, we first give a brief review on three popular 
analytic models that 
explain the $M_{BH,f}- \sigma$ relation. Then we will give a 
detailed description of our model and compare it with other existing 
models.

\section{Review on models of the $M_{BH,f}- \sigma$ relation}
\subsection{Super Eddington Accretion model}
\citet{King2} presented a model to explain the $M_{BH,f}-\sigma$ 
relation. He assumed that the gas density 
profile of a protogalaxy is isothermal ($\rho \sim r^{-2}$) 
\citep{King2,King}. Therefore the gas mass inside radius $R$ is: 
\begin{equation}
M(R)=4 \pi \int_0^R \rho r^2dr= \frac{2f_g \sigma^2 R}{G},
\end{equation}
where $f_g \approx 0.16$ is the cosmological ratio of baryon to total 
mass, assumed to be the same in a galaxy, and the Virial Theorem is used. 
Consider a super-Eddington accretion 
onto a seed blackhole. The accretion feedback produces a momentum-driven 
superbubble that sweeps ambient gas into a thin shell which expands to the 
galaxy. The equation of motion is
\begin{equation}
\frac{d}{dt}[M(R) \dot{R}]+ 
\frac{GM(R)[M_{BH}(t)+M(R)]}{R^2}=\frac{L_{edd}}{c},
\end{equation}
where $L_{edd}=4 \pi G M_{BH}(t)c/ \kappa$, with $\kappa$ the
opacity and $M_{BH}(t)$ the mass of the central blackhole at time $t$. 
Integrating 
twice and assuming $R \gg 
GM_{BH,f}/ \sigma^2$, one gets:
\begin{equation}
R^2=R_0^2+2R_0 \dot{R}_0t- \sigma^2 \left(1- \frac{M_{BH}(t)}{M_{\sigma}} 
\right)t^2,
\end{equation}
where $\dot{R}_0=\dot{R}$ at $R=R_0$, with $R_0$ some large radius ($\gg 
GM_{BH,f}/ \sigma ^2$), and 
$M_{\sigma} \equiv f_g \kappa \sigma^4/ 
\pi G^2$.
Therefore, the maximum radius $R_{\rm max}$ is given by 
\begin{equation}
\frac{R_{\rm max}^2}{R_0^2}=1+ \frac{\dot{R}_0^2}{2 
\sigma^2(1-M_{BH}(t)/M_{\sigma})}.
\end{equation}
When $M_{BH}(t)$ approaches $M_{\sigma}$, $R_{\rm max}$ becomes very large 
such that the cooling of the shocked wind 
is inefficient as the cooling time $t_{cooling} \propto R^2$ and the 
accretion is stopped
as the shell can escape the 
galaxy entirely by gas pressure \citep{King2}. Therefore, given an 
adequate 
mass supply (such as in a merger), we get \citep{King}
\begin{equation}
M_{BH,f}= \frac{f_g \kappa}{ \pi G^2} \sigma^4.
\end{equation}
Here, the proportionality constant $f_g \kappa/ \pi G^2$ lies within the 
observational constraints. To summarize, the $M_{BH,f}- \sigma$ relation 
is 
obtained with three important assumptions: (1) isothermal 
gas density 
distribution throughout the galaxy formation, (2) super 
Eddington accretion, and (3) an adequate mass supply.

\subsection{Self-similar model}
\citet{MacMillan} obtained a relation between $M_{BH,f}$ and 
$\sigma$ 
by assuming that the density and velocity distributions of matter are 
self-similar. They assumed 
that the galaxy 
is formed by the extended collapse of a halo composed of collisionless 
matter. The central blackhole is grown proportionally to the halo as 
matter continues to fall in. The relation is 
given by \citep{MacMillan}
\begin{equation}
\log M_{BH,f} \propto \left( \frac{3 \delta / \alpha -2}{ \delta/ \alpha 
-1} \right) \log \sigma,
\end{equation}
where $\delta$ and $\alpha$ are scales in space and time respectively, and 
their ratio is related to the power-law index of the initial density 
perturbation $\epsilon$ in the spherical infall model of halo growth 
\citep{Henriksen}:
\begin{equation}
\frac{\delta}{\alpha}= \frac{2}{3} \left(1+ \frac{1}{\epsilon} \right).
\end{equation}
The power-law index $\epsilon=(n+3)/2$, where $n$ is the index of the 
primordial matter power spectrum $P(k) \propto k^n$. If 
$n=-2$, Eq.~(8) 
agrees with the observation $M_{BH,f} \propto \sigma^4$. This 
model involves a relation Eq.~(9) which is quite model dependent.

\subsection{Ballistic model}
\citet{Adams} assume the dark matter and baryons to be
unsegregated and 
the isothermal initial mass density distribution ($M_t(r) \propto r$). The 
specific orbital energy is 
conserved when the particles fall into the small seed blackhole:
\begin{equation}
E= \frac{1}{2}v_r^2+ \frac{j^2}{2r^2}- \frac{GM_t(r)}{r},
\end{equation}
where $v_r$ and $j$ are the radial velocity and angular momentum per unit 
mass. When the particles fall into the equatorial plane, 
their pericenters are 
\begin{equation}
p= \frac{j^2}{2GM_t(r)}=\frac{(GM_t(r))^3 \Omega^2}{2 \sigma^8},
\end{equation}
where $\Omega$ is the average angular speed of the particles. In the 
early
stage, all the particles fall into the blackhole until the blackhole mass 
reaches a critical point that corresponds to $p=4R_s$, where 
$M_{BH}(t)=M_t(R_s)$, with 
$R_s=2GM_{BH}(t)/c^2$ the 
Schwarzschild radius. This gives the final relation \citep{Adams}
\begin{equation}
M_{BH,f}= \frac{4 \sigma^4}{Gc \Omega}.
\end{equation}
This model is based on the 
assumption of the isothermal distribution of matter, and 
there is a free parameter $\Omega$, which is assumed to have the same 
value for all galaxies.

\section{Decaying sterile neutrino halo model}
Suppose the sterile neutrino halo dominates the mass in the
proto-galactic center, most 
of the mass in the blackhole comes from the sterile neutrino halo with 
radius $\tilde{R}$, which was formed in the 
very early universe $t \sim 0$ \citep{Munyaneza,Chan}. The total mass of 
the 
degenerate sterile neutrino halo at time $t_b$ is
\begin{equation}
M_s(t_b)=4 \pi \int_0^{\tilde{R}} \rho_sr^2dr=M_{s0}e^{-\Gamma_{3 \nu} 
t_b},
\end{equation}
where $\rho_s$ is the mass density of the sterile 
neutrino halo. Assume that a seed blackhole with mass $M_{BH,0}$ of order 
solar mass is formed at $t_b$, long after the formation of the sterile 
neutrino halo. It would grow by 
accreting mass of the sterile neutrino halo to mass 
$M_{BH}(t)$ at time $t$. As some sterile neutrinos would be accreted by 
the seed 
blackhole, the degenerate pressure is decreased and more sterile neutrinos 
will fall into the blackhole as their Fermi speed is less than their 
escape speed \citep{Munyaneza}. The falling time scale at distance $r$ 
from the blackhole in a free falling model is given by 
\citep{Phillips}
\begin{equation}
t_{ff}=\frac{\pi}{2} \sqrt{ \frac{r^3}{2G[M_{BH,0}+M_s(r,t)]}}.
\end{equation}
For $M_s(t) \ge 10^6M_{\odot}$, $m_s \ge 10$ keV and $\tilde{R} \le 0.04$ 
pc, $t_{ff} \le 160$ 
years for $r \le \tilde{R}$, which is much shorter than the Hubble time. 
Therefore, we do 
not need any intermediate 
mass blackholes since the small seed blackhole can grow to a $10^6-10^9 
M_{\odot}$ supermassive blackhole rapidly as long as there is enough mass 
in the initial sterile neutrino halo. In the following, we assume that all 
sterile neutrinos fall into the blackhole and decay into active
neutrinos 
and photons, and so 
$M_{BH,f}=M_s(t_b)+M_{BH,0} \approx M_{s0}e^{-\Gamma_{3 \nu} t_b}$.

The photons emitted by the original decaying sterile neutrinos provide 
energy to 
the gas in the protogalaxies by Compton scattering. The optical depth of 
a decayed photon in 
the bulge is $\tau=n_e \sigma_TR_e$, where $R_e$ is the $J$-band 
effective bulge radius \citep{Marconi}, $\sigma_T$ is Compton scattering 
cross section and $n_e$ is the mean number density 
of the gas. In equilibrium, the heating 
rate is equal to the cooling rate by Bremsstrahlung radiation $\Lambda_B$, 
Recombination $\Lambda_R$ 
and adiabatic expansion $\Lambda_a$. We have \citep{Katz}
\begin{equation}
L(1-e^{-\tau})= \Lambda_B+ 
\Lambda_R+ \Lambda_a= \left[ \Lambda_{B0}n_e^2T^{0.5}+ 
\Lambda_{R0}n_e^2T^{0.3} \left(1+ \frac{T}{10^6~{\rm K}} 
\right)^{-1} \right]V+PV^{2/3}c_s,
\end{equation}
where $\Lambda_{B0}=1.4 \times 10^{-27}$ erg s$^{-1}$, $\Lambda_{R0}=3.5 
\times 10^{-26}$ erg s$^{-1}$, $c_s$, $P$ and $V$ are the sound speed,
pressure and total volume 
of the gas within $R_e$ respectively. The $M_{BH,f}- \sigma$ relation can 
be obtained by using Eq.~(15) and the Virial theorem numerically. 
Nevertheless, we first illustrate the idea by obtaining analytic 
relations in two different 
regimes. Suppose $M_s(t_b) \ge 
10^6M_{\odot}$; if $\tau \gg 1$, the resulting temperature is above $10^6$ 
K and the total cooling rate is dominated by $\Lambda_a$. For $\tau \le 
1$, the resulting temperature is lower and the total cooling rate is 
dominated by $\Lambda_B$ and $\Lambda_R$ (see Fig.~1). 

In the optically thick regime, $\tau \gg 1$ and $\Lambda_a \gg \Lambda_B+ 
\Lambda_R$, and we get
\begin{equation}
kT= \left( \frac{m_g}{\gamma} \right)^{1/3}V^{-4/9}L^{2/3}n_e^{-2/3},
\end{equation}
where $m_g$ is the mean mass of a gas particle. By using the Virial 
theorem $kT=f_1GM_Bm_g/3R_e$, where $M_B$ is the effective bulge mass of 
the protogalaxy within $R_e$, and substituting $L \approx M_s \Gamma 
c^2/2$, we get
\begin{equation}
M_s= \frac{2 \gamma^{1/2}f_1^{3/2}G^{3/2}e^{-\Gamma_{3 \nu} t}}{3^{7/6}(4 
\pi)^{1/3} 
\Gamma 
c^2} \left( \frac{M_B}{R_e} \right)^{5/2},
\end{equation}
where $\gamma$ is the adiabatic index of the gas and $f_1 \sim 1$ is a 
constant that 
depends on the density distribution 
of the protogalaxy. As time 
passes, the energy gained by the gas would decrease gradually and the mass 
distribution at the center would change slightly also. If the 
supermassive blackhole was formed when the galaxy formation was nearly 
completed ($t_b=10^{16}-10^{17}$ s), the ratio $M_B/R_e$ and the velocity 
dispersion do not change significantly. According to 
Eq.~(17), the ratio $M_B/R_e$ is fixed by $M_s$ and $\Gamma$. 
By using the Virial theorem again and assuming spherical symmetry, 
one can relate this ratio with the final bulge velocity dispersion 
after the supermassive blackhole is formed,
\begin{equation}
\sigma^2=f_2 \frac{GM_B}{R_e},
\end{equation}
where $f_2 \sim 1$ is a constant that depends on the mass distribution at 
present. Combining Eq.~(17) and Eq.~(18), we get
\begin{equation}
M_{BH,f}=M_s(t_b)= \frac{2 \gamma^{1/2}f_1^{3/2}e^{-\Gamma_{3 \nu} 
t_b}}{3^{7/6}(4 \pi)^{1/3}f_2^{5/2}G \Gamma c^2} \sigma^5.
\end{equation} 

In the optically thin regime, $\tau \le 1$ and $\Lambda_R+ \Lambda_B \gg 
\Lambda_a$, the total power absorbed by the gas in the protogalaxies 
within $R_e$ is $\approx Ln_e 
\sigma_TR_e$. If the cooling rate is dominated by Bremsstrahlung 
radiation, in equilibrium, we get 
\begin{equation}
kT=k \left( \frac{L \sigma_TR_e}{\Lambda_0n_eV} \right)^2.
\end{equation}
By using the Virial theorem and Eq.~(18), we obtain
\begin{equation}
M_{BH,f}= \frac{2 \Lambda_{B0}e^{- \Gamma_{3 \nu} t_b}}{\sigma_Tf_2^{3/2}G
\Gamma c^2} \left( 
\frac{f_1}{3km_g} \right)^{1/2} \sigma^3.
\end{equation}
If the cooling rate is dominated by recombination, in equilibrium and for 
$T \le 10^6$ K, we get
\begin{equation}
kT=k \left( \frac{L \sigma_TR_e}{\Lambda_0n_eV} \right)^{10/3},
\end{equation}
and
\begin{equation}
M_{BH,f}= \frac{2 \Lambda_{R0}e^{-\Gamma_{3 \nu} t_b}}{m_g^{7/10} 
\sigma_Tf_2^{13/10}G \Gamma c^2} 
\left( \frac{f_1}{3km_g} \right)^{3/10} \sigma^{2.6}.
\end{equation}
Therefore, $M_{BH,f}$ and $\sigma$ are closely related in both optically 
thick and thin regimes. 

We use 500 
random data in the ranges of $\tau=0.01-10000$, 
$f_1=0.6-3$, $f_2=0.6-3$, $t_b=10^{16}-10^{17}$ s,
$10^9M_{\odot} \le M_B \le 10^{12}M_{\odot}$ and $0.1$ kpc $\le R_e \le 
10$ kpc to generate the values of $M_{BH,f}$ and $\sigma$ by using 
Eq.~(15) and (18) (see 
Fig.~2). By using $\Gamma_{3 \nu}=(5 \pm 1) \times 10^{-17}$ 
s$^{-1}$ which solves the cooling flow problem \citep{Chan}
\footnote{If 
we consider
the major decaying channel is $\nu_s \rightarrow 3 \nu$, the decay rate
obtained in this paper should correspond to $\Gamma_{3 \nu}$.} and 
$\Gamma_{3 \nu}=128 \Gamma$, we get 
\begin{equation}
\log \left( \frac{M_{BH,f}}{M_{\odot}} \right)=(3.98 \pm 0.29) \log \left( 
\frac{\sigma}{200~ \rm km~s^{-1}} \right)+(8.04 \pm 0.20),
\end{equation}
which agrees with the recent observation: $\alpha=4.02 \pm 0.32$ and 
$\beta=8.13 \pm 0.06$ \citep{Tremaine}.

As an order of magnitude estimate, 
$\sin \theta \sim m_D/m_s$, where $m_D$ is the neutrino Dirac mass. In the 
standard see-saw mechanism, $m_D$ is the geometric mean of the light 
neutrino 
mass-scale $m_{\nu}$ and $m_s$ \citep{Mohapatra}. Therefore, $\sin \theta 
\sim 
\sqrt{m_{\nu}/m_s}$. From Eq.~(1), we get
\begin{equation}
\Gamma_{3 \nu} \sim 10^{-23} \left( \frac{m_{\nu}}{\rm 1~eV} \right) 
\left( \frac{m_s}{\rm 1~keV} \right)^4 ~{\rm s^{-1}}.
\end{equation}
For $\Gamma_{3 \nu} \sim 10^{-17}$ s$^{-1}$, $m_s \sim 30$ keV for 
$m_{\nu} \sim 1$ eV, 
which is consistent with our assumption ($m_s \ge 10$ keV).

\section{Discussion and summary}
We assume the existence of a degenerate neutrino halo ($m_s \sim$ keV) at 
the 
proto-galactic center with a parameter, $\Gamma_{3 \nu}$, which is 
universal and
can be inferred from observation of cluster hot gas \citep{Chan}. 
Without any assumptions of the protogalaxy and existence of any 
intermediate mass blackohles, $\alpha \approx 4$ and $\beta \approx 8$ 
require the total decay rate to be $\Gamma_{3 \nu}=(5 \pm 1) \times
10^{-17}$ 
s$^{-1}$, which is 
consistent with the observational data from cooling flow clusters 
\citep{Chan}. Also, 
we have reviewed several models to explain the $M_{BH,f}-\sigma$ 
relation. 
These models require several assumptions or free parameters which may not 
be true 
for all galaxies. For example, King's model assumes the 
isothermal density profile ($\rho \sim 
r^{-2}$) for all galaxies during all the time of the blackhole 
formation. If the density profile changes into the 
form
$\rho \sim r^{-1}$, then $\sigma \propto \sqrt{r}$ which is not 
a constant. This problem also exists in the Ballistic model 
which is based on the isothermal distribution of matter. On the other 
hand, in the self-similar model, there are several free 
parameters which are model dependent. Also one cannot obtain 
the proportionality constant $\beta$ of the $M_{BH,f}- \sigma$ 
relation. 

We have 
considered a wide range of $\tau=0.1-10000$, $f_1=0.6-3$ and $f_2=0.6-3$ 
encompassing almost all possibilities in galaxies. In our model, we assume 
that the supermassive blackhole was formed at the epoch when the galaxy 
formation was nearly completed ($t_b=10^{16}-10^{17}$ s) so that 
the 
velocity dispersion does not change significantly between $t_b$ and 
present. 
Therefore our result is valid only for supermassive blackholes formed 
nearly at the end of the galaxy formation, same as in the 
Super Eddington Accretion and Ballistic model. The assumed 
existence of a decaying sterile neutrino halo inside each galactic center
provides enough mass to form the supermassive blackhole. It can also 
solve the cooling flow problem in clusters \citep{Chan} and explain
the reionization of the 
universe \citep{Hansen}, all with the same decay rate $\Gamma_{3 \nu}=(5 
\pm 1) \times 10^{-17}$ s$^{-1}$ and $m_s \ge 10$ keV, which are 
consistent with the standard see-saw mechanism.

\section{Acknowledgement}
This work is partially supported by a grant from the Research Grant 
Council of the Hong Kong Special Administrative Region, China (Project No. 
400805).

\vskip 10mm

\begin{figure*}
\vskip5mm
 \includegraphics[width=84mm]{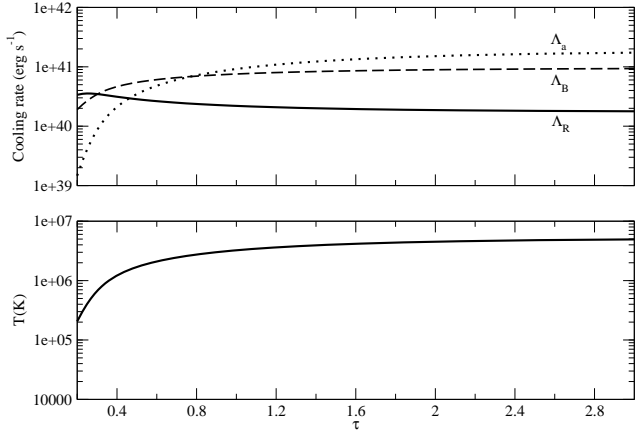}
 \caption{Top: The cooling rates by Bremsstrahlung radiation (dashed 
line), recombination (solid line) and adiabatic expansion (dotted line) 
versus $\tau$ in Eq.~(15). Bottom: The temperature of the gas in 
a protogalaxy versus $\tau$. We used $n_e=1$ cm$^{-3}$, $R_e=1$ kpc and 
$L=3 \times 10^{43}$ erg~s$^{-1}$.} 
\end{figure*}

\begin{figure*}
\vskip5mm
 \includegraphics[width=84mm]{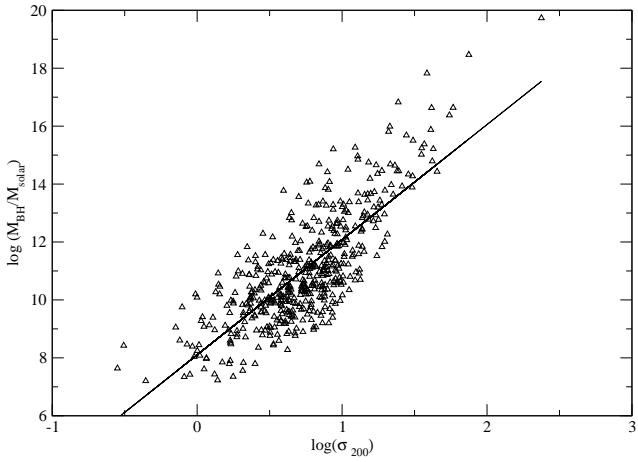}
 \caption{$\log M_{BH,f}/M_{\odot}$ versus $\log \sigma_{200}$ for 500 
random data, where $\sigma_{200}= \sigma /200 \rm km~ 
s^{-1}$. We used $\tau=0.01-10000$, $f_1=0.6-3$, 
$f_2=0.6-3$, $t_b=10^{16}-10^{17}$ s$^{-1}$, 
$M_B=10^9-10^{12}M_{\odot}$ and $R_e=0.1-10$ kpc. The best-fit line in the 
figure corresponds to slope $\alpha=3.97 \pm 0.14$ and intercept 
$\beta=8.11 \pm 0.12$.}
\end{figure*}

\end{document}